\documentclass[twocolumn,prl,aps,showpacs,superscriptaddress]{revtex4}
\usepackage{epsfig}
\usepackage{bm,hyperref}
\usepackage{amssymb}
\def\R{\bm{R}}

\begin{document}    
\title{Adiabatic Theory of Nonlinear Evolution of Quantum States}
\author{Jie Liu}
\affiliation{Department of Physics, The University of Texas, Austin,
Texas 78712}
\author{Biao Wu}
\affiliation{Department of Physics, The University of Texas, Austin,
Texas 78712}
\affiliation{Solid State Division, Oak Ridge National
Laboratory, Oak Ridge, Tennessee 37831-6032}
\author{Qian Niu}
\affiliation{Department of Physics, The University of Texas, Austin,
Texas 78712}                

\date{August 1st, 2002}

\begin{abstract}
We present a general theory for adiabatic evolution of quantum states as
governed by the nonlinear Schr\"odinger equation, and provide examples of
applications with a nonlinear tunneling model for Bose-Einstein condensates.
Our theory not only spells out conditions for adiabatic evolution of
eigenstates, but also characterizes the motion of non-eigenstates which
cannot be obtained from the former in the absence  of the superposition
principle. We find that in the adiabatic evolution of non-eigenstates,
the Aharonov-Anandan phases play the role of classical canonical actions.
\end{abstract}
\maketitle 
Adiabatic evolution has been an important method of preparation and 
control of quantum states\cite{bergmann,raizen,comp}. The main guidance 
comes from the adiabatic theorem of quantum mechanics
\cite{book}, which dictates that an initial nondegenerate eigenstate 
remains to be an instantaneous eigenstate when the Hamiltonian changes slowly 
compared to the level spacings. More precisely, the
quantum eigenstate evolves only in its phase, given 
by the time integral of the eigenenergy (known as the dynamical phase) 
and a quantity independent of the time duration (known as the geometric 
phase).  The linearity of quantum mechanics then immediately allows 
a precise statement about the adiabatic evolution of non-eigenstates through 
the superposition principle.

Our concern here is how the adiabatic theorem gets modified in
nonlinear evolution of quantum states.  Nonlinearity has been introduced in
various forms as possible modifications of quantum mechanics on the 
fundamental level\cite{weinberg}. Our motivation, however, derives 
from practical applications in current pursuits of adiabatic control 
of Bose Einstein condensates (BECs)\cite{bec}, which can often be
accurately described by the nonlinear Schr\"odinger equation. Here the
nonlinearity stems from a mean field treatment of the interactions 
between atoms. Difficulties in theoretical study of adiabatic control 
of the condensate arise not only from the lack of unitarity in the 
evolution of the states but also from the absence of the superposition 
principle.  This problem was recently addressed for BECs in some 
specific cases\cite{band}, and a similar problem was 
discussed in the past for soliton dynamics\cite{kivshar}.

In this Letter, we present a general adiabatic theory for the nonlinear 
evolution of quantum states (eigenstates or noneigenstates).
Our study is conducted by transforming the nonlinear Schr\"odinger equation 
into a mathematically equivalent classical Hamiltonian problem, where 
nonlinearity is no longer a peculiar issue but rather a common character. 
We can thus apply the adiabatic theory for classical 
systems\cite{arnold,LL83,landau} to the study of adiabatic evolution of
quantum states.  The eigenstates become fixed points in the classical problem,
whose adiabatic evolution can be understood from a stability analysis of 
such points.  Aharonov-Anandan phases\cite{aa} 
of cyclic or quasi-cyclic quantum states play the role of canonical action 
in the classical problem, and are therefore conserved during adiabatic 
changes of the control parameters. In the linear case, i.e.,
standard quantum mechanics, these conserved actions are just the
occupation
probabilities on the eigenenergy levels and the quantum adiabatic
theorem becomes a special case of classical adiabatic theorem.
Our results are illustrated with
a nonlinear two-level model\cite{wuniunlz}, which describes
the tunneling of BECs. 
 
{\bf General formalism ---} The quantum system we consider
follows the nonlinear Schr\"odinger equation, 
\begin{equation}\label{eq:nls}
i{d|\Psi(t)\rangle\over d t}=H(|\Psi\rangle,\langle\Psi|,\R)|\Psi(t)\rangle\,,
\end{equation}
where $\hbar$ has been set to one and $\R$ denotes all the system
parameters subject to adiabatic change.
For clarity and simplicity, we focus in this Letter on quantum systems
of finite number of levels, that is,
$|\Psi\rangle={\rm col}(\psi_1,\psi_2,\cdots,\psi_N)$
and 
$\langle\Psi|=(\psi_1^*,\psi_2^*,\cdots,\psi_N^*)$
.
We also assume that the system is invariant under the global 
phase transformation since it is the case of most interest in physics.
The total probability $\langle\Psi|\Psi\rangle$ is then conserved and is 
set to unity hereafter. 
As shown below, it is possible to separate out the 
dynamics of the 
overall phase, and to describe the motion of the remaining variables exactly 
in terms of an $(N-1)$ dimensional classical Hamiltonian dynamics.

First, we write $|\Psi\rangle=e^{i \lambda(t)}|\Phi\rangle$, where the overall phase 
$\lambda$ may be chosen to be the phase of any component, say, $\psi_N$.
Time evolution of the overall phase is given in terms of $|\Phi\rangle$ as derived 
from Eq.(\ref{eq:nls})
\begin{equation}\label{eq:lambda}
\frac{d\lambda(t)}{dt} =
\langle\Phi(t)|i\frac{\partial}{\partial
t}|\Phi(t)\rangle
-\langle\Phi(t)|H|\Phi(t)\rangle\,.
\end{equation}
The first term is related to
a geometric phase, called the Aharonov-Anandan phase \cite{aa}, 
which will be a key component of our theory.

Second, without the total phase $\lambda$, 
the state $|\Phi\rangle$ becomes a member of the so-called projective Hilbert space 
and has $2(N-1)$ independent real variables, which can be conveniently chosen as 
\begin{eqnarray}
{\bm Q}&=&(|\Phi_1|^2, |\Phi_2|^2, \cdots, |\Phi_{N-1}|^2),\\
{\bm P}&=&(\arg(\Phi_1), \arg(\Phi_2), \cdots, \arg(\Phi_{N-1})). 
\end{eqnarray}
They turn out to form a canonical set for an effective classical Hamiltonian 
dynamics as shown below.

Finally, the dynamics of these variables may be obtained from the time-dependent 
variational principle of the original nonlinear Schr\"odinger 
equation (\ref{eq:nls}) with the Lagrangian 
\begin{eqnarray}\label{eq:lag}
\mathcal{L}&=&\langle\Psi|i{\partial\over \partial t}|\Psi\rangle
-\mathcal{H}(|\Psi\rangle,\langle\Psi|,\R)\,\nonumber\\
&=&
\langle\Phi|i{\partial\over\partial t}|\Phi\rangle
-\mathcal{H}(|\Phi\rangle,\langle\Phi|,\R) - {d\over dt} \lambda(t)\,,
\end{eqnarray}
where $\mathcal{H}$ is the total energy of the system and we have used the 
gauge symmetry of the Hamiltonian with respect to the overall phase.
Note that $\mathcal{H}=\langle\Phi|H|\Phi\rangle$ is true only for
a linear system; for a nonlinear system, we generally have 
$H|\Phi\rangle=\delta\mathcal{H}/\delta(\langle\Phi|)$.
Because the overall phase enters in the Lagrangian only through a term of 
total time derivative, the dynamics of the $2(N-1)$ independent variables 
of $|\Phi\rangle$ are closed.  
With the generalized coordinates introduced in Eqs.(3,4) and 
the relation $\langle\Phi|i{\partial\over\partial t}|\Phi\rangle
=-{\bm Q}\cdot\frac{d{\bm P}}{dt}$, in virtue of  the variational
principle of
the Lagrangian, we find that motions of the generalized coordinates are 
governed by  a classical Hamiltonian 
${\mathcal H_{cl}}({\bm Q},{\bm P},\R)=
{\mathcal H}(|\Phi({\bm Q},{\bm P})\rangle,
\langle\Phi({\bm Q},{\bm P})|,\R)$, with the
equations of motion
\begin{equation}\label{eq:motion}
\frac{d {\bm  Q}}{dt} = \frac{\partial \mathcal{H}_{cl}}{\partial
{\bm P}},~~~    
\frac{d {\bm P}}{dt} = -\frac{\partial \mathcal{H}_{cl}}{\partial{\bm Q}}.
\end{equation} 
In this way, we have cast the nonlinear quantum problem (\ref{eq:nls})
into a formalism of classical dynamics, which will allow us 
to use some important results on the adiabaticity of classical mechanics. 

{\bf Eigenstates ---} We first consider adiabatic evolution of 
quantum eigenstates, which can be defined as extremum states of the system energy by 
$H(|\Psi^j(\R)\rangle,\langle\Psi^j(\R)|,\R)|\Psi^j(\R)\rangle 
=E^j(\R)|\Psi^j(\R)\rangle ,
$ 
where the eigenenergies are Lagrange multipliers for
 imposing the normalization condition of the states.  
The eigenstates correspond to fixed points in
 the classical dynamics (6) at a given $\R$\cite{fixed}.  
For an elliptic fixed point, we expect it to be able 
to follow adiabatically 
the control parameter provided the latter changes slowly 
compared with the fundamental frequencies of 
periodic orbits around the fixed point. The frequencies 
can be evaluated by linearizing Eq.(\ref{eq:motion}) 
about the fixed point \cite{arnold,LL83,landau}.
In standard linear quantum mechanics, these frequencies are 
just the level spacings, so that breakdown of adiabaticity occurs
 by level crossing. In the nonlinear quantum problem,
 the fundamental frequencies are in general 
different from the level spacings, so that adiabaticity
 can often be maintained even if the energy levels 
cross as demonstrated later with  a  two-level model.

Nonlinearity in our quantum problem not only makes
different eigenstates non-orthogonal to 
each other, but also can produce more eigenstates than the 
dimension $N$ of the Hilbert space.  Some of 
these additional eigenstates correspond to hyperbolic points
 in the classical dynamics,
characterized by complex fundamental frequencies and strong sensitivity
 to small perturbations.  
We thus expect that such quantum eigenstates not to be able to follow 
adiabatically the control parameter $\R$. In addition, 
quantum states corresponding to elliptic fixed points
can collide and annihilate with a hyperbolic point
during a change of the parameters,
leading to the breakdown of adiabaticity.

\begin{figure}[!htb]
\begin{center}
\includegraphics[width=7.6cm]{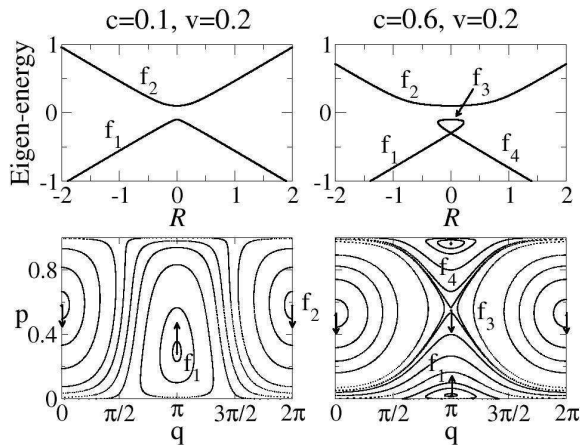}
\end{center}
\caption{The two top panels show eigen-energies as a function of
$R$ for two typical cases $c<v$ and $c>v$. The two bottom
panels show the corresponding phase space portraits
at a given value of $R$(=-0.05). The arrows on the fixed points
indicate the directions of their movements as $R$ increases.}
\label{fig:pd}
\end{figure}

As an illustration, we consider a nonlinear two-level model,
\begin{equation}\label{eq:nlz}
i{\partial\over\partial t}|\Psi\rangle
=\Big({R\over 2}\sigma_z-{c\over 2}\langle\Psi|\sigma_z|\Psi\rangle\,\sigma_z+
{v\over 2}\sigma_x\Big)|\Psi\rangle\,,
\end{equation}
where $|\Psi\rangle={\rm col}(a,b)$ and $\sigma_{z,x}$ is Pauli matrix.
This model was proposed to describe the tunneling 
of Bose-Einstein condensates in optical lattices\cite{wuniunlz,zobay}
or in a double-well potential\cite{dw}. In the model,
the parameter $c$ characterizes the interaction strength
between atoms; $v$ is the coupling strength between the two modes.
The parameter $R$ can be the Bloch wave number\cite{wuniunlz} 
or energy difference between the two wells \cite{dw}.
We are interested in the tunneling between the energy levels
shown in the top panels of Fig.\ref{fig:pd} when $R$ is increased
slowly from the far negative end to the far positive end.

Following the scheme in our general theory,
we choose the total phase as $\lambda=\varphi_b=\arg(b)$ and 
introduce a pair of canonical variables,
$q=\varphi_a-\varphi_b$ and $p=|a|^2$.
The total energy of the system is
$\mathcal H
=\frac{v}{2}\langle\Psi|\sigma_x|\Psi\rangle
+\frac{R}{2}\langle\Psi|\sigma_z|\Psi\rangle 
-\frac{c}{4}\langle\Psi|\sigma_z|\Psi\rangle\langle\Psi|\sigma_z|\Psi\rangle.$
 Then we have the equivalent
classical Hamiltonian as in Eq.(\ref{eq:motion}),
\begin{equation}\label{eq:nlzpq}
\mathcal H_{cl} =v\sqrt{p(1-p)}\cos(q)+{R\over2}(2p-1)
-{c\over 4}(2p-1)^2\,.
\end{equation}

Fig.\ref{fig:pd} shows in the top panels
the structure of eigenenergy levels of Eq.(\ref{eq:nlz})
while the phase space orbits of the corresponding classical 
system (\ref{eq:nlzpq}) is shown in the bottom panels.
When $c<v$, there are only two eigenstates
and two fixed points. Both fixed points are 
surrounded by periodic orbits and hence elliptic. 
The frequencies of the surrounding orbits approach
the fundamental frequencies of the fixed points as they
get closer to  the fixed points. Since both $f_1$ and $f_2$ are
elliptic, the corresponding quantum states are expected to
be able to follow the adiabatically changing $R$.
This has been corroborated by our numerical simulations.

When $c>v$, the situation becomes very different. 
In the energy band, there are two more eigenstates,
forming  a loop structure; in the phase space portrait, there appear two 
more fixed points with one of them, $f_3$, being hyperbolic.
Due to this structure change, the adiabatic evolution becomes
very different here. First, the eigenstate corresponding to $f_3$
will not be able to follow the adiabatic change of $R$ since
$f_3$ is hyperbolic. This has been  checked  by our
numerical integration of Eq.(\ref{eq:nlz}). 
Second, even elliptic point may not able to follow adiabatically.
The fixed point $f_1$ can annihilate itself by colliding
with $f_3$ as $R$ changes slowly, leading to the breakdown
of adiabaticity of the tunneling as reported numerically
in Ref.\cite{wuniunlz}. We also notice that there is
a level crossing between $f_1$ and $f_4$ at $R=0$; however, 
our calculation shows that their fundamental frequencies
are $v((\frac cv)^2-1)^{1/2}$, not equal to zero.
This clearly illustrates our statement  
in the general theory that the fundamental frequencies are not
related to the level spacing in the nonlinear case.

{\bf Cyclic and quasi-cyclic states} 
Compared with eigenstates, adiabatic evolution of non-eigenstates is in 
general very complicated as 
the motions given by Eq. (\ref{eq:motion}) may be chaotic.  We choose to focus 
on the quantum states around an elliptic point, where the classical orbits 
are regular.  In particular, for the nonlinear 
two-level problem, the classical dynamics is completely integrable.  
The non-eigenstates in such cases are 
cyclic or quasicyclic states, in which the system returns or almost returns to 
its original state after an evolution.  On the $(N-1)$-dimensional torus of 
the regular region, we may introduce a set  of action-angle variables,
with ${\bm I}=(I_1, I_2, ..., I_{N-1}),
{\bm \Theta}=(\Theta_1, \Theta_2,...
\Theta_{N-1})$\cite{arnold,LL83,landau}.
The angular variables change with time with the
frequencies ${\bm \omega}=
(\omega_1,\omega_2,...,\omega_{N-1})$ while the actions
${\bm I}$ are constants.  More importantly,
according to the classical adiabatic theorem \cite{LL83,landau}, 
the actions ${\bm I}$ are adiabatic invariants in the sense that they remain 
constant even if the control parameter $\R$ changes (slowly) in time.  The existence 
of these adiabatic invariants presents strong constraint on the motion, 
and guarantees a quantum state initially close to an 
eigenstate (elliptic point) to stay near it as the system is changed adiabatically.

Interestingly, we can attach a physical meaning to these adiabatic invariants in the 
effective classical description by making connection to the Aharonov-Anandan (AA) phase of 
the quantum states.  The AA phase is defined as the time integral
of the first term in Eq.(\ref{eq:lambda}) for a periodic orbit or a 
quasi-periodic state\cite{add}, 
\begin{equation}\label{eq:aa}
\gamma_{AA}({\bm R}) =
\int_0^\tau \langle\Phi(t)|i\frac{\partial}
{\partial t}|\Phi(t)\rangle dt
\end{equation}
We can rewrite it with the canonical variables $({\bm Q},{\bm P})$ and further
with the action-angle variables,
\begin{equation}\label{eq:action}
\gamma_{AA}= \oint {\bm P}d{\bm Q}={\bm I}\cdot{\bm \Omega}\,.
\end{equation}
where ${\bm \Omega}=(\omega_1\tau,\omega_2\tau,\cdots,\omega_{N-1}\tau)$
and $\tau$ is time period.
Therefore, 
the actions are related to the AA phase $\gamma_{AA}$, which is an
observable physical quantity \cite{exppp}. In the special case of $N=2$,
there is only one independent action,
 so the AA phase is simply $\gamma_{AA}=2\pi I$. This
simple connection can be expanded to the general case of $N>2$, where
one can single out a particular cyclic state that involves with only one
action $I_n$. For this cyclic state, we again have the simple relation 
$\gamma_{AAn}=2\pi I_n$. As a result, one can identify
these AA phases $\gamma_{AAn}$ as the adiabatic invariants in place of
the actions $I_n$.

How do the above adiabatic invariants connect to the familiar
notions in the standard linear quantum mechanics? 
Consider the time evolution of a general state 
in a linear quantum system
for a given $\R$
, $|\Psi\rangle=\sum_n c_ne^{iE^nt}|n\rangle$.
where $E^n$'s are the eigenenergies. This is a (quasi-)cyclic state with 
the projective wave function given by
$|\Phi\rangle=\sum_{n=1}^{N-1}c_ne^{i(E^n-E^N)t}|n\rangle+c_N|N\rangle$.
Its AA phase can be computed with Eq.(\ref{eq:aa}); after comparing
with Eq.(\ref{eq:action}), we immediately find that $I_n=|c_n|^2$.
Therefore, in the linear quantum mechanics, these adiabatically invariant 
actions $I_n$ are nothing but the probabilities on the energy levels. 
In this way, we have rederived the adiabatic theorem of linear quantum 
mechanics which states that the probability on each energy 
level is conserved in adiabatic processes. Note that this derivation of the 
quantum adiabatic theorem through our effective classical description 
is distinct from the usual semiclassical
 relation  discussed in Ref.\cite{mberry}.

\begin{figure}[!htb]
\begin{center}
\includegraphics[width=5cm]{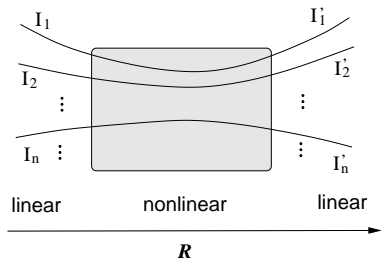}
\end{center}
\caption{Nonlinear tunneling of a system that is
nonlinear in an intermediate range of the parameter $\R$.
$I$'s and $I^\prime$'s are the occupation
probabilities on different eigenstates, at the beginning and
the end of the tunneling process, respectively.}
\label{fig:ntun}
\end{figure}
The conservation of probabilities on the energy levels 
can be generalized to the case where the system 
is nonlinear only in an intermediate range of the parameter $\R$
(see Fig.\ref{fig:ntun}). When the actions (or AA phases) are conserved 
during the entire process, the initial and final probabilities 
on the energy levels must remain the same, $I_n=I_n^{\prime}$,
because the system is linear at the beginning and end of 
the process so the actions are just the occupation probabilities.
On the other hand, in the intermediate range where the nonlinearity dominates,
the probabilities will change (even oscillate greatly) since the 
conserved actions
are not probabilities on the energy levels. 
\vspace{0.5cm}
\begin{figure}[!htb]
\begin{center}
\includegraphics[width=8cm]{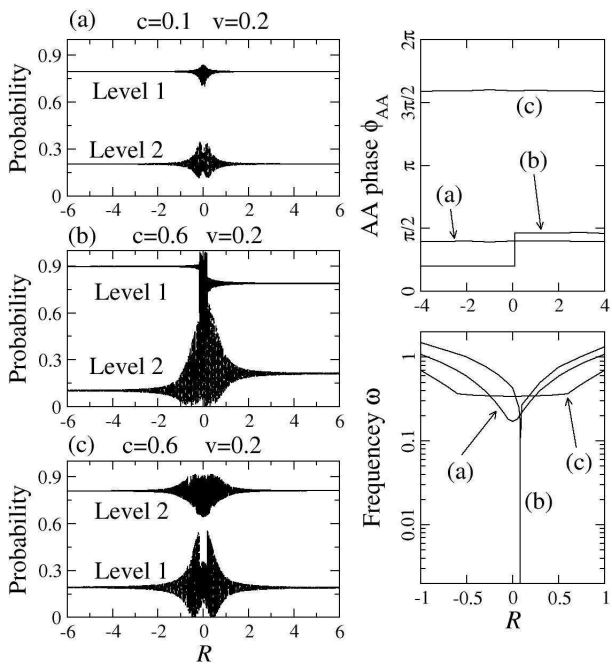}
\end{center}
\caption{Left panels: Change of probabilities on the two levels with
$R$, which changes with rate $\alpha=0.0001$, for 
three different cases. The right two 
panels show how the AA phases and the fundamental frequencies 
change with $R$ in these three cases, respectively.}
\label{fig:pol}
\end{figure}

We now illustrate this important result using our two-level model,
where, at the two infinite ends with $|R|\gg c$, 
the nonlinear term can be ignored and the system
is effectively linear. For $c<v$, where all the 
fixed points are elliptic, the fundamental frequency $\omega$ for the 
periodic orbit remains finite and the AA phase (action) is conserved (see
lines (a) in the right panels of Fig.\ref{fig:pol}).  
The initial and final probabilities on each level 
are indeed the same (Fig.\ref{fig:pol}(a)), 
although they oscillate in the intermediate range of the parameter
where the system is nonlinear. 
As the nonlinearity gets strong, the occurrence of tunneling
begins to depend on the choice of the initial state.
In case (b) of Fig.\ref{fig:pol}, where one starts
with probability $I=0.1$ on level two, tunneling happens;
however, in case (c) where one starts with
probability $I=0.8$ on level two, there is no tunneling.
The difference is whether there is collision with the
hyperbolic point $f_3$. In case (b), the initial noneigenstate falls on
a periodic orbit surrounding the fixed point $f_1$, which will later
collide with the hyperbolic point $f_3$, where the fundamental frequency 
drops to zero and the AA phase has a finite jump 
(see lines (b) in the right panels). The jump height is proportional to
the tunneling probability.
In case (c), the initial state falls on
a periodic orbit around the fixed point $f_2$, which 
will not collide with $f_3$. 

In conclusion, we have generalized the standard quantum adiabatic
theorem to nonlinear quantum systems by transforming them into
mathematically equivalent classical Hamiltonians.
In the classical systems, the eigenstates become fixed points
and their adiabatic evolutions are determined by whether these
fixed points are elliptic or hyperbolic. Furthermore, we have
found that the adiabatic evolutions of non-eigenstates are
controlled by AA phases, which play the role of classical actions.
Within the same framework we have also considered the 
case of closed-loop parameter change as pioneered by Berry\cite{berry}, we 
find that Hannay's angles\cite{hannay} are accumulated at the return of 
the quantum state (details will be published in the future).

We acknowledge the support of NSF, R.A. Welch Foundation, 
and the International Center of Quantum Structures in Beijing.


\begin{thebibliography}{99}
\bibitem{bergmann}K. Bergmann {\it et al.},
Rev. Mod. Phys. {\bf 70}, 1003 (1998).

\bibitem{raizen} M.B. Dahan {\it et al.}, Phys. Rev. Lett. 
{\bf 76}, 4508 (1996); S.R. Wilkinson {\it et al.}, 
Phys. Rev. Lett. {\bf 76}, 4512 (1996).

\bibitem{comp} S.Das, {\it et al} Phys. Rev. A {bf 65}, 062310 (2002);
N.F.Bell, {\it et al}  Phys. Rev. A {\bf 65}, 042328 (2002);
A.M.Childs, {\it et al} Phys. Rev. A {65}, 012322 (2002); R.G.Unanyan,
{\it et
al}
Phys. Rev. Lett. {\bf 87}, 137902 (2001).

\bibitem{book} L.D. Landau and E.M. Lifshitz, {\it Quantum Mechanics}
(Pergamon Press, New York, 1977).

\bibitem{weinberg}S. Weinberg, Phys. Rev. Lett. {\bf 62}, 485 (1989);
S. Weinberg, Ann. Phys. {\bf 194}, 336 (1989).

\bibitem{bec}F. Dalfovo {\it et al.}, 
Rev. Mod. Phys. {\bf 71}, 463 (1999);
A.J. Leggett, Rev. Mod. Phys. {\bf 73}, 307 (2001);
R. Dum {\it et al.}, Phys. Rev. Lett. {\bf 80}, 2972 (1998);
Z.P. Karkuszewski, K. Sacha, and J. Zakrzewski, Phys. Rev. A{\bf 63},
061601(R) (2001);
T. L. Gustavson,{\it et al},
Phys. Rev. Lett. 88, 020401 (2002);
J. Williams, {\it at al} 
Phys. Rev. A 61, 033612 (2000);
Matt Mackie, {\it et al} 
Phys. Rev. Lett. 84, 3803 (2000);   
Roberto B. Diener, Biao Wu, Mark G. Raizen, and Qian Niu,
Phys. Rev. Lett. {\bf 89}, 070401 (2002).

\bibitem{band}
Y.B. Band, B. Malomed, and M. Trippenbach, 
Phys. Rev. A {\bf 65}, 033607 (2002); Y.B. Band and M. Trippenbach, 
Phys. Rev. A {\bf 65}, 053602 (2002); G.J. de Valc\'{a}rcel,
cond-mat/0204406.

\bibitem{kivshar}Y.S. Kivshar and B.A. Malomed, Rev. Mod. Phys. {\bf 61},
763 (1989).


\bibitem{arnold} 
V.I. Arnol'd, {\it Mathematical Methods of Classical Mechanics} 
(Springer-Verlag, New York, 1978).

\bibitem{LL83}A.J. Lichtenberg and M.A. Lieberman, {\it Regular and
Stochastic Motion} (Springer-Verlag, 1983);
W. Dittrich and M. Reuter, {\it Classical and Quantum Dynamics:
from classical paths to path integrals} (Springer-Verlag, 1992).

\bibitem{landau} 
L.D. Landau and E.M. Lifshitz, {\it  Mechanics}
(Pergamon Press, New York, 1976).        

\bibitem{aa}For quantum mechanics, see 
Y. Aharonov and J. Anandan, Phys. Rev. Lett. {\bf 58},
1593, (1987). For nonlinear systems, see
J.C. Garrison and R.Y. Chiao, \prl {\bf 60}, 165 (1988);
J. Anandan, Phys. Rev. Lett. {\bf 60}, 2555 (1988).

\bibitem{wuniunlz}Biao Wu and Qian Niu, Phys. Rev. A {\bf 61}, 
023402 (2000).


\bibitem{fixed}For an eigenstate $|\Psi^j\rangle$, its time evolution involves
only change of phase, $|\Psi^j\rangle(t)=e^{i\phi(t)}|\Psi^j\rangle(0)$. With
the erasion of the overall phase $\lambda$, the corresponding state 
$|\Phi^j\rangle$ in the projective space does not change with time, and thus is
a fixed point.

\bibitem{zobay}O. Zobay and B.M. Garraway, Phys. Rev. A {\bf 61},
033603 (2000).

\bibitem{dw}G.J. Milburn {\it et al.}, Phys. Rev. A {\bf 55}, 4318 (1997);
A. Smerzi {\it et al.}, Phys. Rev. Lett. {\bf 79}, 4950 (1997); S. Raghavan
{\it et al.}, Phys. Rev. A {\bf 59}, 620 (1999); I. Marino {\it et al.},
Phys. Rev. A {\bf 60}, 487 (1999);
A.Vardi and J.R.Anglin, Phys. Rev. Lett. {\bf 86} 568 (2001);
I.Zapata, F.Sols and A.J.Leggett, Phys. Rev. A {\bf 57}, R28 (1998);
S. Kohler and F.Sols 
Phys. Rev. Lett. 89, 060403 (2002)   
.


\bibitem{add}
For quasiperiodic motions, where the frequencies $\omega_i$ are not 
commensurate with each other, one can use 
rational numbers $n/m$ to approach the ratios $\omega_i/\omega_j$. 
The accuracy of the approximation can be improved indefinitely with more complex rationals.
In this spirit, Eq.(\ref{eq:action}) also holds for quasi-cyclic motions.

\bibitem{exppp}
D. Suter, K.T. Mueller, and A. Pines, Phys. Rev. Lett. {\bf 60},1218 (1988);
H. Weinfurter, Phys. Rev. Lett. {\bf 64} 1318 (1990).

\bibitem{mberry}M.V. Berry, J. Phys. A: {\bf 17} 1225 (1984).

\bibitem{berry}
M.V. Berry, Proc. R. Soc. London A {\bf 392}, 45 (1984).

\bibitem{hannay}
J.H. Hannay, J. Phys. A {\bf 18}, 221 (1985);
M.V. Berry and J.H. Hannay, J. Phys. A. {\bf 21}, L325 (1988).


\end{thebibliography}
\end{document}